\documentclass[preprint,12pt]{elsarticle}

\journal{Chemical Physics Letters}

\usepackage{gensymb}
\usepackage{graphicx}
\usepackage{dcolumn}
\usepackage{amssymb}
\newcolumntype{R}{>{\raggedleft\arraybackslash}X}
\usepackage{bm}

\begin{document}

\title{On the relation between hydrogen bonds, tetrahedral order and molecular mobility in model water}

\author{Rodolfo Guillermo Pereyra}
\author{Aleida Berm\'udez di Lorenzo}
\address{Fa.M.A.F., Universidad Nacional de C\'{o}rdoba, C\'{o}rdoba, Argentina and IFEG-CONICET, Medina Allende s/n, Ciudad Universitaria, X5000HUA C\'{o}rdoba, Argentina} 
\author{David C. Malaspina}
\author{Marcelo A. Carignano}
\ead{cari@northwestern.edu}
\address{Department of Biomedical Engineering and Chemistry of Life Processes Institute, Northwestern University, 2145 Sheridan Road, Evanston, IL 60208}


\begin{abstract}
We studied by molecular dynamics simulations the relation existing between the lifetime of hydrogen bonds,  the tetrahedral order and the diffusion coefficient of model water. We tested four different models: SPC/E, TIP4P-Ew, TIP5P-Ew and Six-site, these last two having sites explicitly resembling the water lone pairs. While all the models perform reasonably well at ambient conditions, their behavior is significantly different for temperatures below 270 K. The models with explicit lone-pairs have a longer hydrogen bond lifetime, a better tetrahedral order and a smaller diffusion coefficient than the models without them.
\end{abstract}

\maketitle

\newpage

\section{Introduction}

Water has been the object of a significant effort from the simulation and modeling community\cite{Finney:2001aa,Carignano:2005aa,Vrbka:2006aa,Moore:2010aa}. The problem of accurately representing water with an atomistic model is, in view of the never decreasing number of attempts, a very difficult one. Even though there have been significant advances in the recent years\cite{Guillot:2002aa,Vega:2011aa}, still there is no model that can claim to capture all the essential properties of water. It is therefore important to explore, with as much detail as possible, the behavior of different models under different conditions. For example, it is important to understand what are the effects of the different design features of the models and how different model parameters affect the properties of the simulated bulk system. In this letter we focus our attention on four different models, all having different molecular architecture. In particular, we consider models with a different number of interacting sites, from three to six. 
Varying the number of interacting sites impose a modification of the point charges in order to keep the molecular dipole moment within reasonable values.
Although the four tested models provide a reasonable description of water at ambient conditions, their behaviors diverge as the temperature decreases, as it will be shown below. None of these models provide a description of water that quantitatively corresponds to real water in a temperature range that includes ambient temperature and the supercooled regime. However, it is very important to have a comparative picture of all them, especially because numerous publications aimed to understand properties of water a low temperature are sometimes based on one particular model.\cite{Sciortino:1996aa,Starr:2000aa,Pereyra:2009aa,Malaspina:2010aa}
In this paper we aim to shed light on the effect of the molecular architecture, in particular the effect of explicitly including sites resembling the molecular lone pair orbitals, on the resulting bulk properties of the system.
Our findings suggest that different behaviors are likely to emerge if different water models are used. Within this spirit, we test the four models in a relatively wide range of temperatures; from 230 K to 300 K. This temperature range was purposely chosen to contain the melting temperature and the temperature of spontaneous nucleation.\cite{Debenedetti:2003aa}

\section{Models and Methods}

The models that we have selected for our study are the SPC/E\cite{BERENDSEN:1987aa}, TIP4P-Ew\cite{Horn:2004aa}, TIP5P-Ew\cite{Rick:2004aa} and Six-Site\cite{Nada:2003aa} models. The two TIP models are reparametrizations of their original versions\cite{JORGENSEN:1985aa,Mahoney:2000aa}. An important difference between the TIP5P-Ew and Six-Site with respect to the other two models of this study is the presence of sites explicitly resembling the water lone pairs. One may guess, based on geometrical and energetic considerations, that a model having lone-pairs should have a higher tendency towards tetrahedral molecular arrangements than those models without explicit lone-pairs. This conjecture is supported by our results presented below. 

We performed molecular dynamics simulations in the NPT ensemble. For that, we used the GROMACS v.4.5.5 simulation package\cite{Hess:2008aa}. For all the cases, we used a time step of 1 fs, temperature coupling using the Nos\'e-Hoover thermostat with time constant of 0.1 ps and pressure coupling to a Parrinello-Rahman barostat with a compressibility of $4.5 \times 10^{-5}$ bar$^{-1}$ and time constant of 0.5 ps. The simulated systems contained 512 molecules for the SPC/E, TIP4P-Ew and TIP5P-Ew, and 1536 molecules for the Six-Site models. The long range electrostatic contributions were taken into account by the particle mesh algorithm. A spherical cut-off of 0.9 nm was imposed for the Lennard-Jones terms of the interactions, and for the short range electrostatic contributions. No long range corrections was used for the Six-Site model. The total time of the production runs was increased, from 10 ns to 50 ns as the temperature of the system was decreased. 

Hydrogen bonds were identified using a geometrical definition based on the distance between the oxygen atoms and the 
{O-H} $\cdots$ O angle. We allowed a maximum angle of 30$^\circ$ and a maximum distance equals to the position of the first minimum of the oxygen-oxygen radial distribution function at the corresponding temperatures.  

We define the lifetime of a hydrogen bond ($t_H$) as the time elapsed from the moment of the bond formation until the moment at which the bond breaks. Namely, we use a first breaking method that disregard the possibility for the rapid  bond recovery.\cite{Starr:2000aa} This choice allows for a simple computation of the relation between the hydrogen bond lifetime, and the tetrahedral order parameter (defined below) of the molecules involved in a particular bond. The calculation of $t_H$ requires to monitor the trajectory to find the formation and breaking times. We performed this analysis on the trajectory files saved during the simulation. The frequency at which the trajectory was recorded affects the precision in the determination of $t_H$, so that we need to save the trajectory with a frequency higher than the typical frequency of bond breaking. By performing a series of trials, we determined that a frequency of 1 frame/ps is enough for $T$=300 K. In most of the cases, at every saved frame of the trajectory there are several hydrogen bonds that are broken. By averaging the lifetime of all the bonds that break at time $t$ we calculate the time dependent average lifetime $\overline{t_H}(t)$. The conformation of the system at the start of the analysis contains a number of hydrogen bonds. In order to properly account for the total bond lifetime it is necessary to extend the analysis for a time long enough to allow for all the initial bonds to break. 

\section{Results}
\begin{figure}[!h]
\begin{center}
\includegraphics*[width=0.5\textwidth]{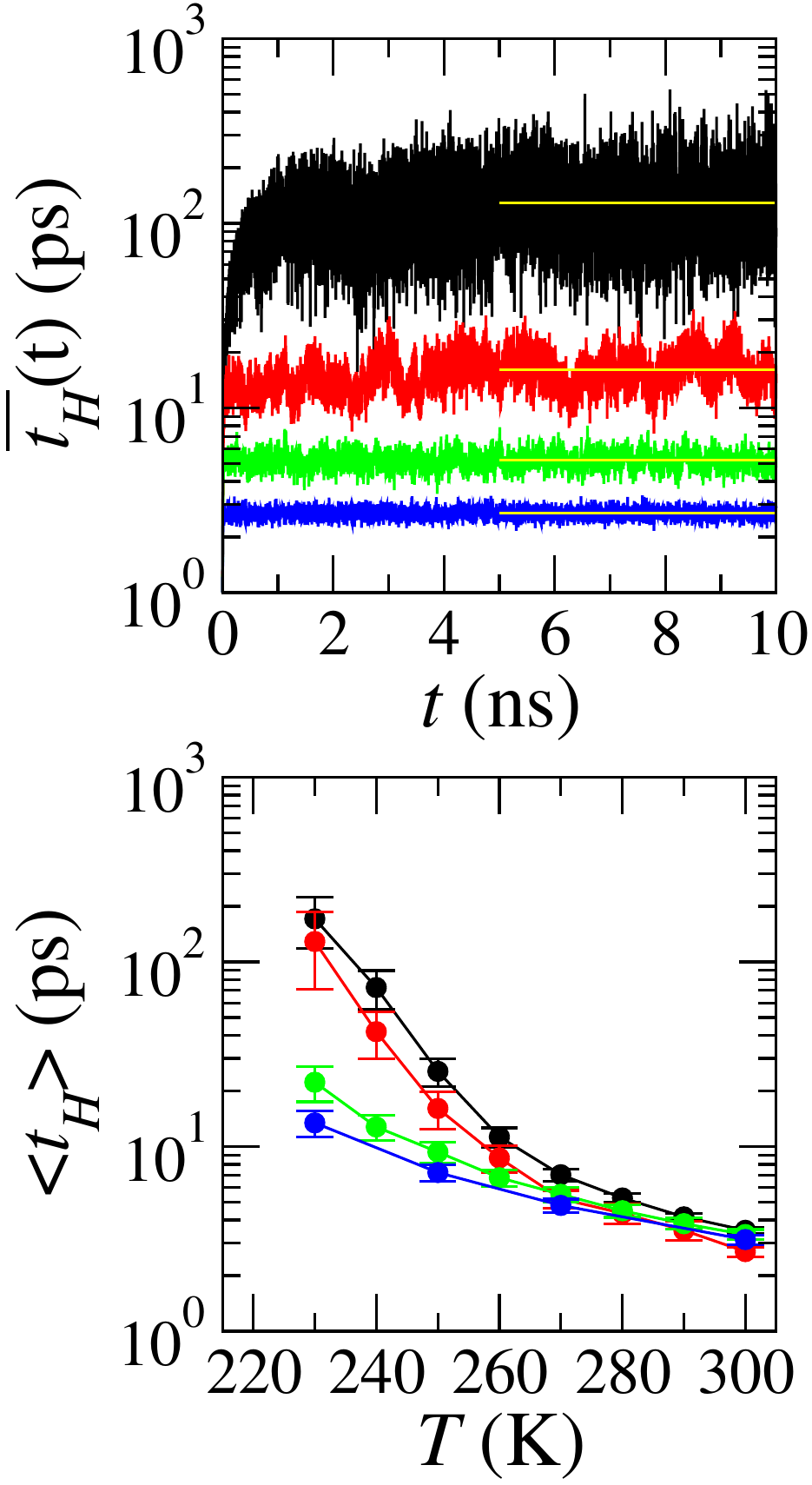}
\end{center}
\caption{Top pannel: System average of the instantaneous hydrogen bond lifetime, $\overline{t_H}(t)$, as a function of time for TIP5P-Ew. Black, red, green and blue correspond to 230 K, 250 K, 270 K and 300 K, respectively. The horizontal lines represent the time average $\langle t_H \rangle$. Bottom pannel: Average hydrogen bond lifetime vs. temperature for Six-Site (black), TIP5P-Ew (red), TIP4P-Ew (green) and SPC/E (blue) models.}
\label{th}
\end{figure}

In Figure \ref{th}a we show an example of the calculation of the hydrogen bond lifetime corresponding to simulations of TIP5P-Ew water at several temperatures. The analysis was performed on the last 10 ns of the production run. The figure shows how $\overline{t_H}(t)$ increases until it reaches a plateau. The time average $\langle t_H \rangle$ on the plateau region, indicated by the horizontal straight lines, represents our estimate for the lifetime of the hydrogen bonds at each temperature. Clearly, $\langle t_H \rangle$ increases as the temperature of the system decreases, reflecting the slower kinetics of the system. Also, it is clear that the dispersion of the data is larger for the lower temperatures. In Figure \ref{th}b we show $\langle t_H \rangle$ vs. $T$ for the four simulated models. The results show that the average hydrogen bond lifetimes are very similar for the four models for temperatures in the range 270 K to 300 K. At lower temperatures, the lifetime increases considerably. However, this increment is more pronounced for the TIP5P-Ew and Six-Site models than for SPC/E and TIP4P-Ew. At 230 K, the hydrogen bond lifetime of the models with explicit lone pairs is an order of magnitude longer than that of the models without lone-pairs.

\begin{figure}[!h]
\begin{center}
\includegraphics*[width=0.65\textwidth]{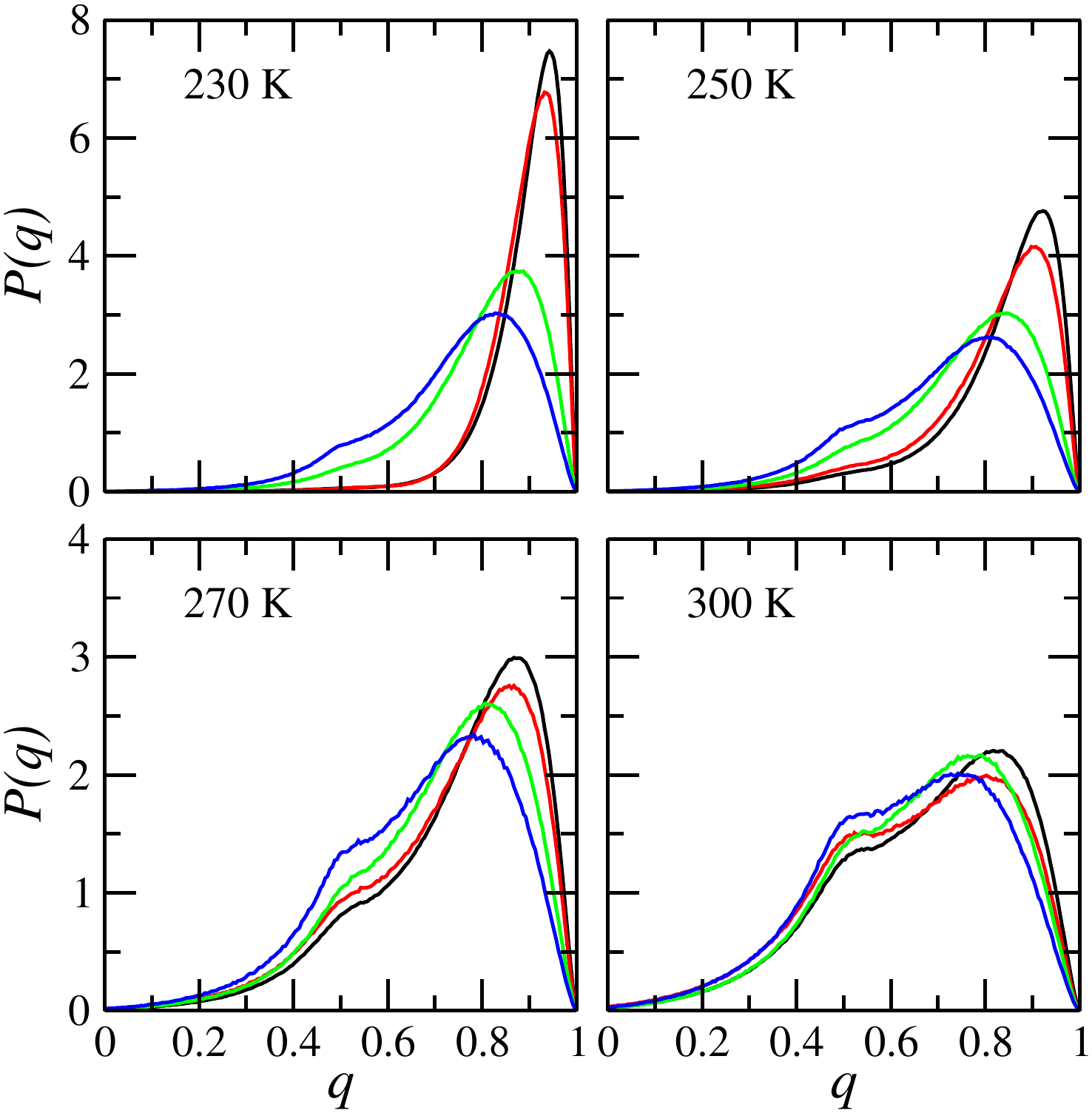}
\end{center}
\caption{Probability density for the tetrahedral order parameter $q$ for the Six-Site (back), TIP5P-Ew (red), TIP4P-Ew (green) and SPC/E (blue) models and different temperatures.}
\label{q}
\end{figure}

\begin{figure}[!h]
\begin{center}
\includegraphics*[width=0.65\textwidth]{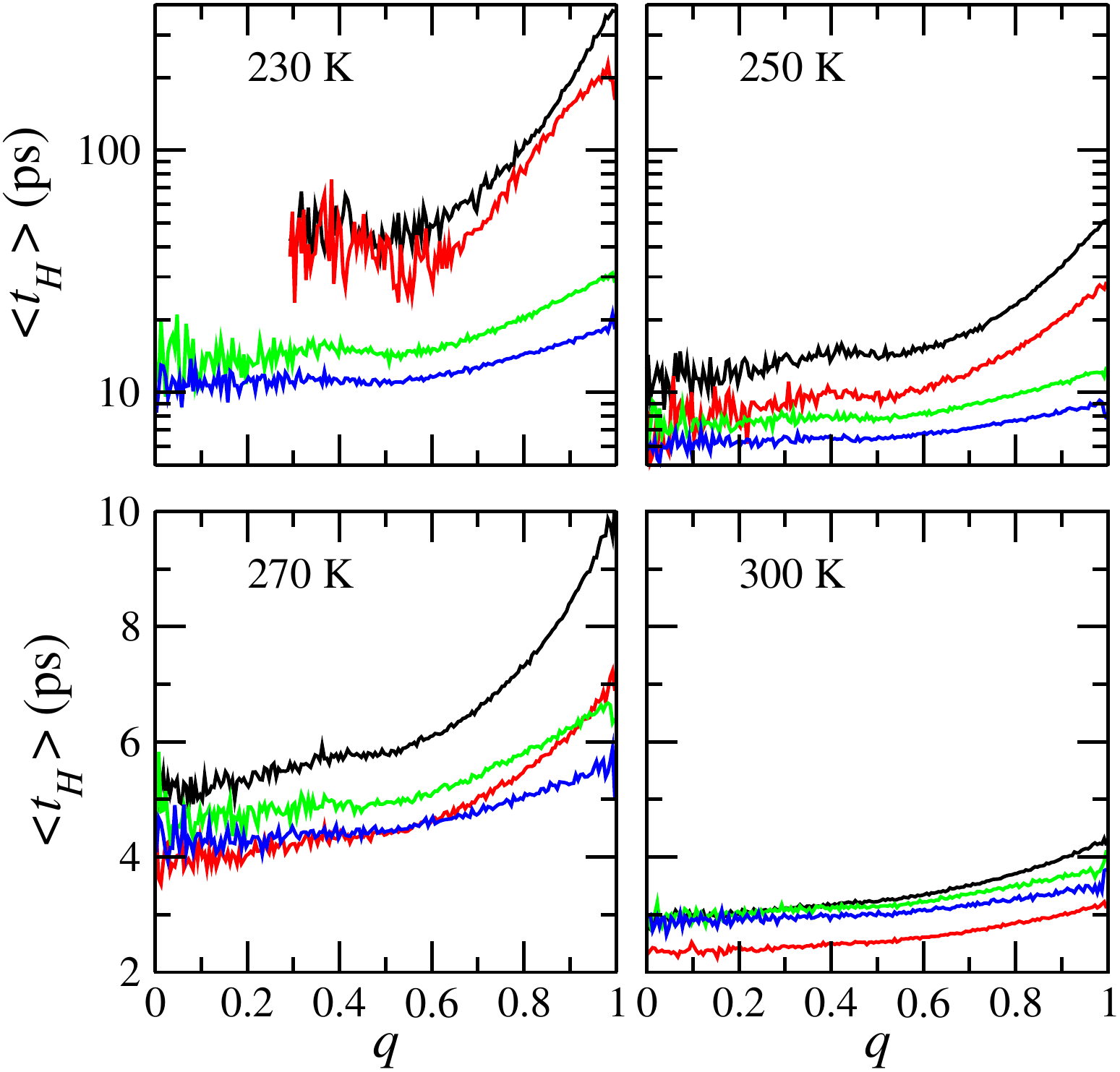}
\end{center}
\caption{Hydrogen bond lifetime vs. tetrahedral order parameter for the Six-Site (black), TIP5P-Ew (red), TIP4P-Ew (green) and SPC/E (blue) models and different temperatures. Note the logarithmic scale used for the lower temperatures, in contrast to the linear scale used for the higher temperatures.}
\label{qth}
\end{figure}

\begin{figure}[!h]
\begin{center}
\includegraphics*[width=0.65\textwidth]{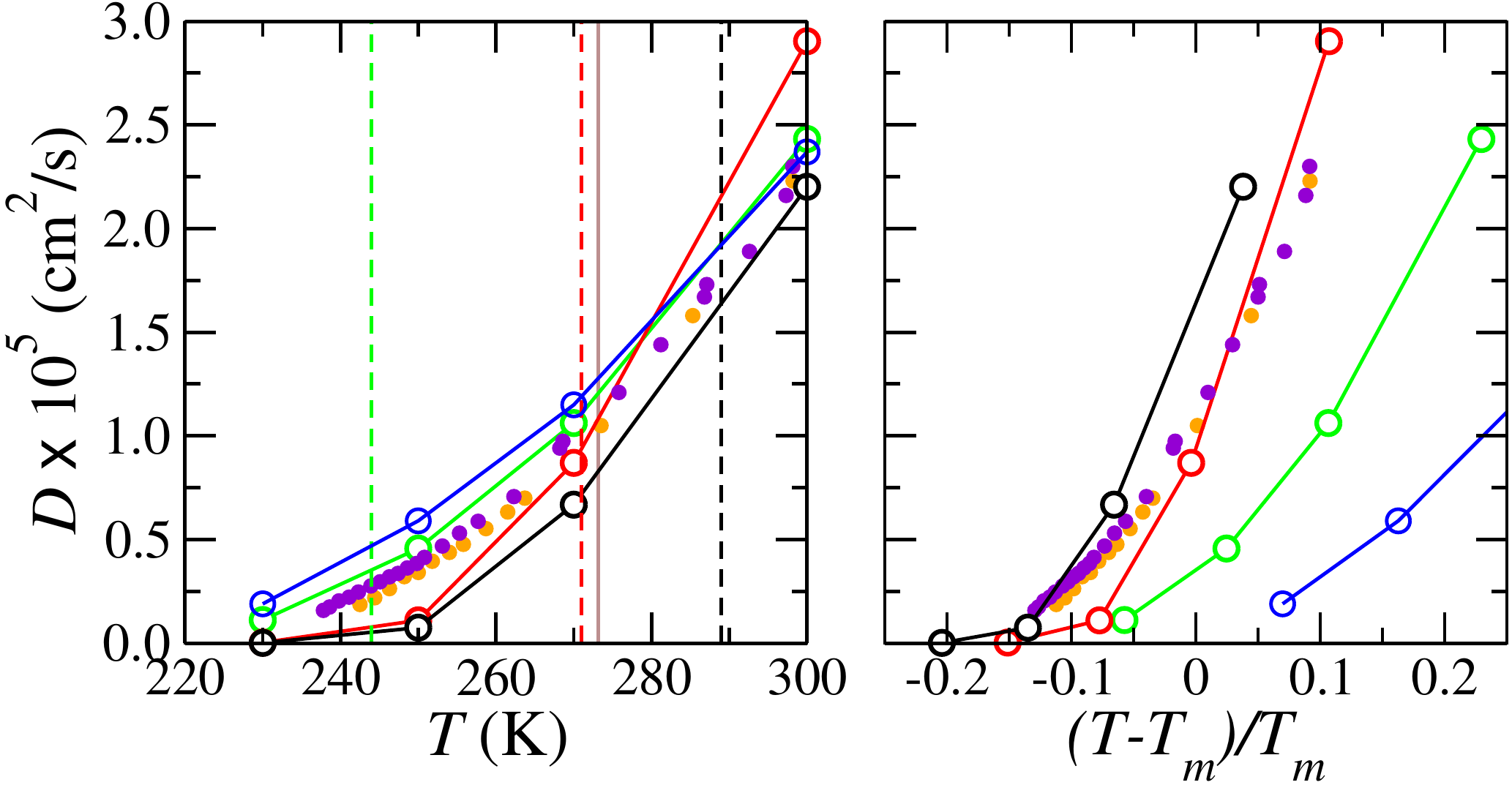}
\end{center}
\caption{Left panel: Diffusion coefficient (open circles) as a function of temperature, and melting temperature (vertical dashed lines) for the four models. Six-Site (black), TIP5P-Ew (red), TIP4P-Ew (green) and SPC/E (blue). The melting temperature for SPC/E falls out of the plotted temperature range. Experimental diffusion coefficients is represented with filled symbols in orange\cite{GILLEN:1972ac} and violet\cite{Price:1999aa}. The vertical brown line indicates the experimental melting temperature. Right panel: Same data plotted as a function of the relative temperature scaled with corresponding melting temperature.}
\label{d1}
\end{figure}

The level of tetrahedral arrangement of molecule $k$ in relation to its four nearest neighbors is  quantified by the local order parameter\cite{Errington:2001aa}
\begin{equation}
q(k)=1-\frac{3}{8} \sum_{i=1}^3 \sum_{j=i+1}^4 \left ({\cos{\psi_{ikj}}+\frac{1}{3}} \right)^2  \,\,\,.
\end{equation}
Here,  $\psi_{ikj}$ is the angle formed by the lines joining the oxygen atom of the central molecule $k$, and those of its nearest neighbors $i$ and $j$. This parameter takes values in the range $-3 \leq q \leq 1$. For perfect tetrahedral order,  $q(k)=1$; and for random molecular order, $\langle q(k) \rangle =0$. In Figure 2  we show the probability density $P(q)$ for the four models at four temperatures. At $T$=300 K, the curves for the four models are remarkably similar and characterized by a bimodal distribution as previously discussed.\cite{Errington:2001aa,Kumar:2009aa} 
This bimodality is also displayed by the local structure index,\cite{Shiratani:1996aa} and has been used in support of the idea that liquid water consists of a mixture of molecules in two different structural states.\cite{Appignanesi:2009aa,Malaspina:2009aa}
As the temperature is decreased, the high-$q$ peak increases at expenses of the low-$q$ peak reflecting the development of a higher tetrahedral order. However, the temperature dependency of this shift to a more structured system is less pronounced for SPC/E and TIP4P-Ew than for the other two models. At $T$=230 K, $P(q)$ for TIP5P-Ew and Six-Site models  show a sharp peak near $q$ = 0.9, while the other two models still show an important contribution at lower $q$. 
In fact, the models without explicit lone pairs show the high-$q$ peak at much lower temperatures. These results, when considered together with the temperature dependency of the hydrogen bonds lifetime, displayed in Figure \ref{th}b, suggest that there is a correlation between $\langle t_H \rangle$ and $q$. This is investigated in Figure \ref{qth} in which we have discriminated the hydrogen bond lifetime by the $q$ value of the molecules involved in those bonds. Since $q$ is an instantaneous value and $\langle t_H \rangle$ is not, we have (arbitrarily) chosen to associate $\langle t_H \rangle$ to the $q$ value at a time just before of the bond breaking. Figure \ref{qth} clearly shows that, for all the temperature range and the four models, those molecules with a higher $q$ participate in hydrogen bonds with longer lifetimes. This is especially evident at the lower temperatures (top pannels) for which we have used a logarithmic scale. For the higher temperatures, the linear scale helps to visualize the same effect, although it is very minor at 300 K. At T=230 K, notice the similarity between the curves corrsponding to the Six-Site and TIP5P-Ew, contrasting with the weaker $q$ dependence of the SPC/E and TIP4P-Ew models. The dispersion observed for $q \lesssim 0.5$ at low temperature is due to the scarcity of the data in this $q$ range, as shown in Figure \ref{q}, and is the reason for which we have suppressed the data for $q < 0.3$. The picture that emerges is that the models with explicit lone pairs have a clear tendency to form more stable and longed lived hydrogen bonds than the model without lone pairs. Since the lifetime of the hydrogen bonds is related to the molecular mobility, for a given temperature it should be expected that the kinetics of SPC/E and TIP4P-Ew is faster than that of the other two models. This is indeed the case, as reflected by the temperature dependence of the diffusion coefficient, shown in Figure \ref{d1}. Interestingly, of the four studied models, the TIP4P-Ew is the model that better approximates the experimental diffusion coefficient in the whole temperature range. Nevertheless, the melting temperature is well approximated by the TIP5P-Ew model ($T_m$=271 K)\cite{Fernandez:2006aa},  underestimated by the SPC/E ($T_m$=215 K)\cite{Vega:2005aa} and TIP4P-Ew ($T_m$=244 K)\cite{Fernandez:2006aa} and overestimated  by the Six-Site ($T_m$=289 K).\cite{Abascal:2006aa} Using the melting temperature as a reference to create a relative reference temperature leaves the TIP5P-Ew as the best approximation but with a steeper temperature dependence than the experimental data.

\begin{figure}[!h]
\begin{center}
\includegraphics*[width=0.75\textwidth]{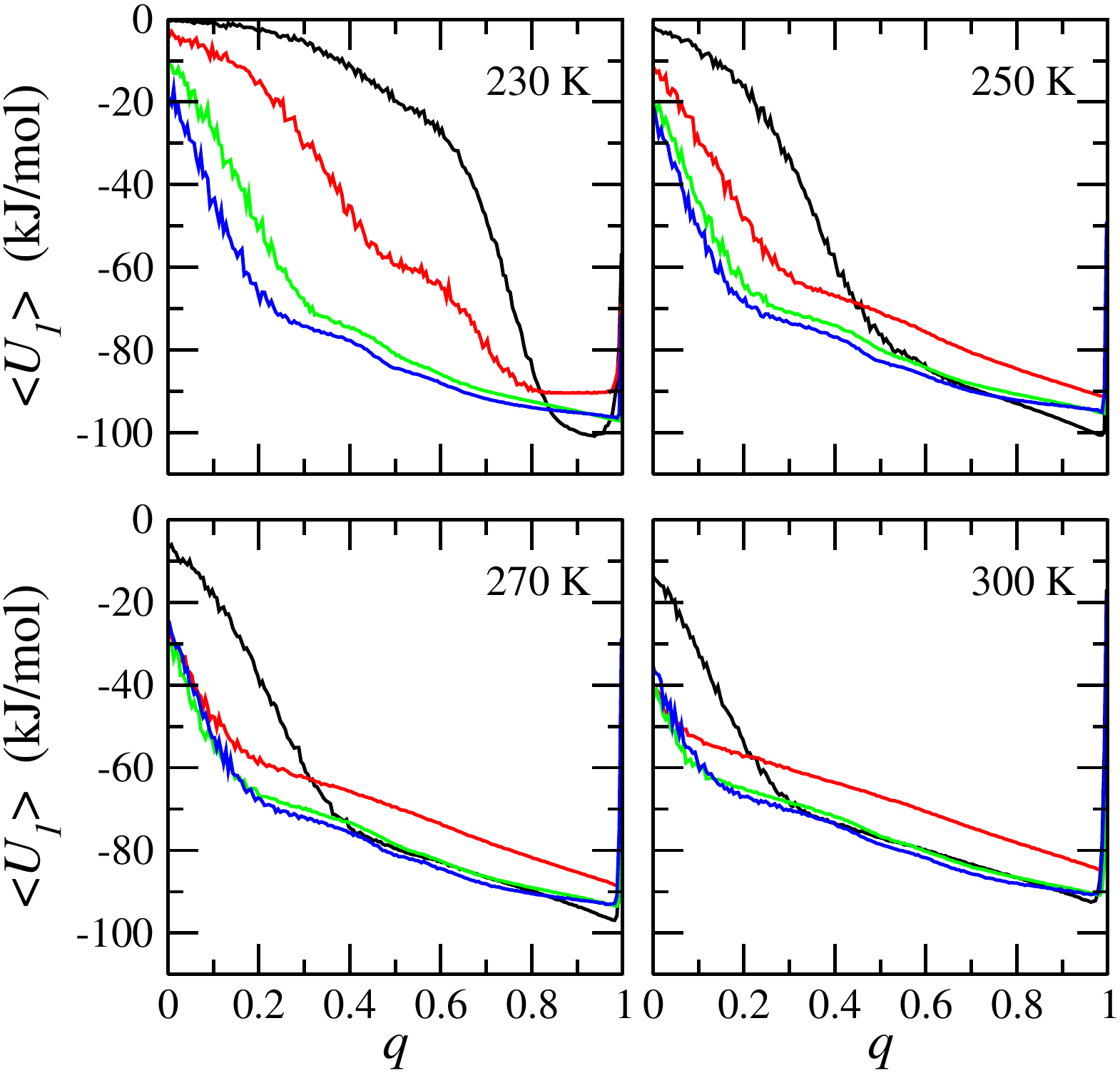}
\end{center}
\caption{Interaction energy of a single molecule with the rest of the system as a function of the tetrahedral order parameter $q$, for the Six-Site (black), TIP5P-Ew (red), TIP4P-Ew (green) and SPC/E (blue) models.}
\label{u1}
\end{figure}

In order to explore further the origin of the relation between explicit lone pairs with higher tetrahedral structure we calculate the dependency of the interaction energy between a single molecule and the rest of the system, with the value of the tetrahedral order parameter $q$ of that molecule. We denote this quantity by $U_1(q)$.  As a reference, the average configurational energy (from 300 K to 230 K) of the four models are: from -41 to -49 kJ/mol for the Six-Site, from -40 to -50 kJ/mol for the TIP5P-Ew, from -46 to -51 kJ/mol for the TIP4P-Ew and from -47 to -51 kJ/mol for the SPC/E, respectively. Note that the magnitude of $U_1(q)$ is larger than the average configurational potential energy, because this last quantity is normalized by the number of molecules in the system. Since we are interested in its qualitative behavior, for this calculation we consider only the short range contribution to the potential energy, i.e, the long range electrostatic corrections are disregarded. The results are presented in Figure \ref{u1}. The curves $U_1(q)$ may be regarded as the average energy landscape experienced by the water molecules around its equilibrium structure in the liquid. At $T$=300 K temperature, the $U_1(q)$ curves suggest that the molecules can explore configurations within a wide range of $q$ values, regardless of the model used to describe the system.  However, the curves do not show a double well at the higher temperatures that will result in the bimodal distribution shown in Figure \ref{q}, possibly due to the thermal fluctuations that blur these energy landscapes. As the temperature is decreased, an energetic penalty develops for those configurations with lower $q$. The curves develop a shoulder, clearly visible for the five and six sites models, around $q \simeq 0.6$. 
For SPC/E and TIP4P-Ew the temperature effect is less dramatic, nevertheless they show a clear restriction for configurations with $q \lesssim 0.3$ and $T$=230 K. Also, for the lowest studied temperature, the TIP5P-Ew and the Six-Site models show clear potential energy wells centered at $q \simeq 0.9$ that are responsible for trapping the system in configurations with high tetrahedral order. The effect is stronger in the Six-Site model than in the TIP5P-Ew, in line with our previous findings, in particular with the probability density $P(q)$ displayed in Figure \ref{q}. 

In conclusion, we have presented evidence suggesting that models having explicit interaction sites resembling the water lone pairs have a clear effect favoring tetrahedral structures. This higher tendency to form tetrahedral structures translate to a higher lifetime of hydrogen bonds and a slower diffusion than the models without explicit lone pairs. Using the melting temperature of the model to rescale the system temperature show that the TIP5P-Ew and Six-Site models are a better approximation to the experimental results than the other two model, and between these two, the TIP5P-Ew appears as the better model since it has a much better approximation to the ice melting temperature.

\section*{Acknowledgments}
Marcelo Carignano acknowledges the support from NSF (grant CHE-0957653).


\pagebreak

\section*{Figures}

\end{document}